# Pure nematic transition inside the superconducting dome of iron chalcogenide superconductor FeSe$_{1-x}$Te$_x$

K. Y. Liang[1,*], R. Z. Zhang[1,*,†], Z. F. Lin[2,3,*], Z. J. Li[1], B. R. Chen[1], P. H. Zhang[1], K. Z. Yao[1], Q. S. He[1], Q. Z. Zhou[1], H. X. Yao[1], K. Jin[2,3,†], Y. H. Wang[1,4,†]

1. State Key Laboratory of Surface Physics and Department of Physics, Fudan University, Shanghai 200433, China
2. Beijing National Laboratory for Condensed Matter Physics, Institute of Physics, Chinese Academy of Sciences, Beijing 100190, China
3. School of Physical Sciences, University of Chinese Academy of Sciences, Beijing 100049, China
4. Shanghai Research Center for Quantum Sciences, Shanghai 201315, China

\* These authors contributed equally to this work.
† Email address: ruozhouzhang@fudan.edu.cn; kuijin@iphy.ac.cn; wangyhv@fudan.edu.cn;

**Abstract**
**Nematicity and magnetism are prevalent orders in high transition temperature ($T_c$) superconductors, coexisting in the parent compound of most material families. Quantum fluctuations of nematicity or spin orders are both plausible candidates for mediating unconventional Cooper pairing. Identifying the sole effect of a nematic quantum critical point (QCP) on the emergence of superconducting dome without interference of spin fluctuations is therefore highly desirable. The iron chalcogenide superconductor FeSe exhibits pure nematicity without any magnetic ordering. A nematic quantum phase transition can be induced by Te substitution but experimental study of such transition is so far limited to its normal state. By performing local susceptometry on composition-spread FeSe$_{1-x}$Te$_x$ films (0 ≤ x ≤ 1) using scanning Superconducting Quantum Interference Device (sSQUID) microscopy, we investigate the superfluid density ($\rho_s$) across the pure nematic transition in extremely fine steps of Δx = 0.0008. The temperature dependence of $\rho_s$ changes from the form of anisotropic pairing on the nematic side to an isotropic one across the critical doping $x_c$. The power-law dependence of gap anisotropy on $|x − x_c|$ provides evidence for nematic quantum criticality under the superconducting dome. The low-temperature $\rho_s$ scales linearly with $T_c$ in the**

**nematic phase $x < x_c$, whereas the gap amplitude, maximized at $x_c$, determines the $T_c$ for $x > x_c$. Our results establish a pure nematic QCP in FeSe$_{1-x}$Te$_x$, separating two superconducting orders with distinct pairing boosted by nematic quantum fluctuations.**

## Introduction

A quantum critical point (QCP) refers to the point where a symmetry-breaking phase of matter is continuously suppressed to zero by tuning a non-thermal parameter at zero temperature. Strong quantum fluctuations of the order parameter exist around a QCP. Symmetry breaking orders of both charge and spin are ubiquitous in the parent compounds of high transition temperature ($T_c$) superconductors. These orders can be gradually suppressed by chemical substitutions for instance, which also leads to the emergence of superconducting dome. The close proximity of the highest $T_c$ point to potential QCP is a tantalizing clue to establish their causal relationship[1,2]. In iron-based superconductors (FeSCs), for example, superconductivity reaches maximum when the electronic nematic order, from a $C_4$ crystalline symmetry to a $C_2$ electronic ordering, is suppressed by chemical doping or pressure to a putative QCP[3–6]. In cuprate superconductors, recent studies also signify nematic QCP near the verge of the pseudogap state[7–9]. It has been theoretically proposed that strong nematic fluctuation near such a QCP may be responsible for superconducting pairing[10–12]. Experimentally elucidating the existence and the role of the nematic QCP beneath the superconducting dome is thus crucial for understanding the microscopic origin of high-$T_c$ superconductivity.

Since the appearance of the superconducting dome obscures a potential QCP of the original symmetry-breaking phase, it is generally difficult to access the QCP to identify the effect of quantum fluctuations on superconductivity. Furthermore, there are typically more than one symmetry breaking order in the parent compound of most high $T_c$ materials. For cuprate superconductors, the coexistence of nematicity with spin density wave poses significant challenges in isolating the influence of nematic transition. Similar situation occurs in FeSCs, where the nematic phase always overlaps with an itinerant antiferromagnetism (AFM) order, making the putative nematic QCP and the AFM QCP indistinguishable (Fig. 1a)[13,14]. In particular, a sharp peak of London penetration $\lambda$ is observed near the critical doping of several iron pnictide superconductors[13,15,16], which is attributed to the enhancement of quasiparticle mass due to the quantum critical fluctuations. However, it is yet unknown whether this anomaly corresponds to a nematic QCP or an AFM QCP.

The iron chalcogenide superconductor FeSe$_{1-x}$Te$_x$ (FST) is a unique system whose nematic phase is decoupled from magnetic ordering at ambient pressure[17]. Isovalent substituting Se with Te monotonically suppresses the nematic order, accompanied by a putative nematic QCP at $x_c$ (around 0.5 for bulk crystals and 0.2-0.4 for thin films), as evidenced by the divergent nematic susceptibility observed in the normal state[18–20] (Fig. 1b). Moreover, the absence of significant spin fluctuations near $x_c$ has been revealed by nuclear magnetic resonance measurements[21,22]. These advantages render FST an ideal material platform to study the nonmagnetic nematic QCP in high-$T_c$ superconductors. However, direct evidence for the nematic QCP inside the superconducting dome is still lacking so far. This is largely because conventional technique to probe normal state nematic criticality such as elastoresistivity is no longer effective in the superconducting state due to vanishing resistance. Applying a strong magnetic field can suppress the superconducting phase to expose the 'bare' nematic QCP, but this may induce a different QCP or shift the zero-field QCP[15,23,24].

One key parameter characterizing the superconducting state is the superfluid density ($\rho_s \equiv \lambda^{-2} \propto n_s/m^*$, where $n_s$ is the Cooper pair density and $m^*$ is the effective mass), which encodes the phase rigidity of the pairing condensate. The temperature ($T$) dependence of $\rho_s$ is associated with the quasiparticle energy spectrum that depends on the superconducting gap structure[25–27]. Importantly, it is found that the gap anisotropy of orthogonal FST is intimately related to its nematic order through various scenarios, e.g., the mixture of $s$- and $d$- wave components in the $A_{1g}$ symmetry in the nematic phase[17], the nematic pairing due to orbital-selective correlations or spin fluctuations in the nematic phase[28–30], pairing in the presence of nematic orbital ordering[31], the strong gap anisotropy due to the shrink of electron pocket below the nematic transition[32,33]. Hence, the temperature dependence of superfluid density may manifest the nematic order inside the superconducting dome.

Here, we perform local $\rho_s$ measurements on composition-spread FST films via scanning superconducting quantum interference device (sSQUID) microscopy. The continuous variation of Te content from 0 to 1 on a single substrate in conjunction with the micron spatial resolution of sSQUID enables us to explore $\rho_s$ across the pure nematic transition in unprecedently fine steps, $\Delta x = 0.0008$. By analyzing the $\rho_s(T)$ data for different compositions, we find the superconducting gap structure of FST undergoes a transition from anisotropic symmetry to isotropic symmetry at $x_c \sim 0.23$. Close to $x_c$, the gap anisotropy obeys a power-law dependence on Te content, evidencing the nematic quantum criticality in the superconducting state. The averaged gap value reaches maximum at $x_c$, strongly supports the scenario that pairing interaction is strengthened by the nematic fluctuations associated with the QCP. On the other hand, we find that the Te content at which $\rho_s$ peaks derivates from $x_c$ as temperature decreases, possibly due to the suppression of phase coherence by the pinned nematic fluctuations beyond the QCP. Combing $T_c$, $\rho_s$ and gap information together, we find that the pure nematic QCP divides the superconducting dome into two regions, where $T_c$ is dominated by the phase rigidity and the pairing potential for $x < x_c$ and $x > x_c$, respectively.

**Results**

**sSQUID characterization of composition-spread FST films**

Composition-spread FST films with thicknesses of $100 \pm 3$ nm were deposited on $10 \times 10$ mm$^2$ (00$l$)-oriented single-crystal CaF$_2$ substrates by combinatorial laser molecular beam epitaxy technique[34,35]. During the deposition, two polycrystalline targets of FeSe and FeTe were ablated alternatively, and a mobile shadow mask was used to control the deposition area and the thickness on the substrate. Consequently, unit-cell-thick linear Te gradient ($x = 0$ to $1$ and $\Delta x \sim 0.12/$mm) on the substrate can be achieved (Fig. 1d), confirmed by the energy dispersive x-ray spectrometer measurements. The epitaxial (00*l*)-oriented growth of the film was verified by the micro-region x-ray diffraction measurements. The films were further patterned into microbridge arrays by photolithography, which were used for local transport characterization and further served as reference points for composition when navigating the nano-SQUID chip across the sample.

The composition-spread FST films were loaded in our home-built sSQUID microscope which can resolve micrometer-scale variation in both magnetization and susceptibility with high magnetic flux sensitivity[36–39]. In the nano-SQUID chip, a 2-μm-diameter pickup coil was integrated into the loop of a two-junction SQUID that converts the magnetic flux ($\Phi$) through the coil into the voltage signal (the inset of Fig. 1c). A local magnetic field was generated by the ac current ($I_F$) in a 4.5-μm-diameter field coil. By demodulating the in-phase component of the voltage signal, the real part of the ac susceptibility ($\chi' = d\Phi/dI_F$) can be obtained. The nano-SQUID chip was mounted on a quartz tuning fork to provide real-time feedback of the chip-sample distance ($z$). The piezoelectric positioners enable us to change the position of the sample relative to the nano-SQUID chip and perform line-by-line scanning. In order to conduct measurements above $T_c$, the nano-SQUID chip was thermally isolated from the variable-temperature stage to maintain a working temperature of 4.60 K.

By scanning the nano-SQUID chip parallel to the surface of the composition-spread film, we can characterize the magnetic responses for small variation of Te composition limited by our spatial resolution of about 2 microns. For each scan area, the sample exhibits uniform superconducting diamagnetism ($\chi' < 0$, indicated by the blue color) on the micron scale (Fig. 1e). The temperature evolution of the susceptibility images further demonstrates the homogeneous superconducting transition (Fig. S1a). These results suggest the high quality of our composition-spread FST films. Notably, the diamagnetism strength ($|\chi'|$) appears to follow a dome-shaped behavior with Te doping: $|\chi'|$ increases as $x$ increases from ~ 0 to ~ 0.38, but decreases with further increase in $x$. The corresponding magnetometry scans can be found in Supplementary Sections 1 and 2.

For a superconducting thin film ($d \ll \lambda$ with $d$ the sample thickness), the ac susceptibility is proportional to the inverse of the Pearl length, which is essentially the sheet superfluid density according to $\Lambda_p^{-1} = \frac{d}{2}\rho_s$. The absolute value of $\Lambda_p^{-1}$ can be obtained by fitting the susceptibility approach curve $\chi'(z)$ to a thin diamagnet model developed by Kogan[40] and Kirtley et al.[41] (Supplementary Section 3). By varying the nano-SQUID position and the sample temperature ($T$), we have obtained temperature dependent $\chi'(z)$ data for 58 compositions (Fig. S3). The white markers in Fig. 1e indicate the corresponding chemical composition of the $\chi'(z)$ data. Note that for the region where $x$ is close to $x_c$ ~ 0.23 (e.g., region $x$ ~ 0.216), the distance between adjacent points is as small as 7 µm, corresponding to composition change of $\Delta x$ = 0.0008 across the nematic phase boundary. It can be seen from some representative $\chi'(z)$ curves (Fig. 2a) that they are well fitted with the model (the gray solid lines). For $x = 0$, the obtained $\Lambda_p(5\ \text{K}) = 16.4 \pm 0.5$ µm yields London penetration depth $\lambda(5\ \text{K}) = 0.9054 \pm 0.06$ µm, in accord with the value of FeSe film ($1.12 \pm 0.04$ µm) measured by the two-coil mutual inductance technique[42]. The $\chi'(z)$ curve at $x = 0.38$ shows the largest diamagnetic amplitude consistent with the susceptibility image (Fig. 1e).

**Temperature dependences of the superfluid density and the pairing gap**

The superfluid density as a function of temperature $\Lambda_p^{-1}(T)$ clearly exhibits different forms dependent upon the compositions (Fig. 2b). A qualitative observation is that $\Lambda_p^{-1}$ is roughly linear with $T$ in the orthogonal phase ($x < x_c$, red and orange data points), but is relatively curved for $x > x_c$ (blue data points), which is evident from the normalized plot of $\widetilde{\Lambda_p^{-1}} = \Lambda_p^{-1}/\Lambda_p^{-1}$ (5 K) versus reduced temperature $\widetilde{T} = T/T_c$ (Fig. 2c). By displaying the $d\widetilde{\Lambda_p^{-1}}/d\widetilde{T}$ data in the $T$ versus $x$ phase diagram, we find it decreases steeply as Te content is increased across $x_c$ (Fig. 2e). Generally, a deep gap minimum (i.e., strong gap anisotropy) would facilitate the formation of thermally excited quasiparticles, leading to a rapid reduction of $\Lambda_p^{-1}$ with increasing $T$. The abrupt change of $d\widetilde{\Lambda_p^{-1}}/d\widetilde{T}$ thus signifies a transition in the gap anisotropy near $x_c$. Coincidentally, the normal-state resistance ($R$) above $T_c$ evolves from a strange-metal behavior ($dR/dT =$ const.) to a convex behavior ($dR/dT$ increases with decreasing $T$) around $x_c$ (Fig. 2e and Supplementary Section 4), accompanied by a sudden change in the Hall coefficient (Figs. S4c and S4d). The switching of both the $\Lambda_p^{-1}(T)$ and the transport properties near $x_c$ suggests that the pure nematic QCP separates two distinct superconducting phases i.e., SC1 for $x < x_c$ and SC2 for $x > x_c$ with different normal states.

To further investigate the evolution of the gap structure with Te doping, we fit the $\Lambda_p^{-1}(T)$ data using the standard expression within London approximation ($\lambda \gg \xi$)[25,43]:

$$\frac{\Lambda_p^{-1}(T)}{\Lambda_p^{-1}(0\text{ K})} = 1 + \frac{1}{\pi}\int_0^{2\pi}\int_{\Delta(T,\varphi)}^{\infty}\left(\frac{\partial f}{\partial E}\right)\frac{EdEd\varphi}{\sqrt{E^2 - \Delta(T,\varphi)^2}}, \quad (1)$$

where $f = [1 + \exp(E/k_B T)]$ is the Fermi function, $\varphi$ is the angle along the Fermi surface, and $\Delta(T,\varphi) = \Delta_0(T)\delta(T/T_c)g(\varphi)$. Here $g(\varphi)$ describes angular dependence of the gap and $\Delta_0 = \langle\Delta(0,\varphi)\rangle_\varphi$ is the zero-temperature average gap value. The temperature dependence of the gap is approximated by $\delta(T/T_c) = \tanh\{1.82[1.018(T_c/T - 1)]^{0.51}\}$ [44]. We consider a two-fold symmetric gap of FeSe[28,45] so that $g(\varphi)$ takes as an extended s-wave form $g(\varphi) = |1 + \sqrt{2}\alpha\cos(2\varphi)|/\sqrt{1+\alpha^2/2}$, where $\alpha$ controls the degree of the gap anisotropy, with gap nodes appearing for $\alpha \geq 1/\sqrt{2}$. Such a symmetry of gap function has also been adopted to analysis the $T$ dependences of the superfluid density[46] and the specific heat[47,48] of FeSe$_{1-x}$Te$_x$ single crystals. In general, one could choose other harmonics, e.g., $\sim \cos(4\varphi)$ underlying the lattice symmetry[49], or include another gap to account for the multiband feature[50,51]. However, these details would not alter qualitatively the extracted gap parameters presented below (see Supplementary Section 5).

Fig. 3a presents the results of the fits of Eq. (1) to $\Lambda_p^{-1}(T)$ data for selected doping levels, with $\Lambda_p^{-1}(0\,\text{K})$, $\Delta_0$, and $\alpha$ as free parameters (see the fitting results for all doping levels in Fig. S6). The corresponding gap functions in the momentum ($\boldsymbol{k}$) space are given in Fig. 3b. The marked gap anisotropy at $x = 0$ is consistent with the strong $\boldsymbol{k}$ dependence of the gap structure of FeSe, revealed by the bulk thermodynamics[17] and the energy spectroscopy measurements[28,45]. Besides, compared to the line nodes of bulk FeSe crystals suggested by several experiments, the nodeless gap inferred here may originate from the lift of the accidental nodes by disorder or twinned nematic domain[17,52]. With Te doping, the gap anisotropy increases first, and then vanishes continuously as $x$ approaches $x_c$ (see also the top panel of Fig. 3c). The isotropic gap symmetry of the tetragonal FST was also observed by the angle-resolved photoemission spectroscopy (ARPES) measurements[53,54].

As we mentioned, the strong anisotropy in $\Delta(\boldsymbol{k})$ of FST is closely tied to its nematicity, and the doping dependence of $\alpha$ reflects the evolution of nematic order below $T_c$. Particularly for the extended-$s$ wave gap symmetry, $\alpha$ is directly related to the C$_2$ symmetry breaking of the gap structure:

$$\alpha = \frac{4\pi}{3} \frac{\langle \Delta_{k_x} \rangle - \langle \Delta_{k_y} \rangle}{\langle \Delta \rangle}, \quad (2)$$

where $\Delta_{k_x}$ and $\Delta_{k_y}$ are the superconducting gaps along $k_x$ and $k_y$ directions respectively. Thereby, $\alpha$ can be naturally identified as the order parameter of the nematicity[5]. Inspecting the doping dependence of $\alpha$ near $x_c$ (see inset of Fig. 3c), we further find it obeys a power-law dependence $\alpha \propto |x - x_c|^\beta$ ($x < x_c$) with $\beta = 0.55 \pm 0.11$ (the red line). The power-law index is in agreement with the anticipated value of $\beta = 0.5$ for the Ising quantum phase transition with total dimensionality D > 4 (D = $d + z$, with $d$ the effective dimensionality of the quantum fluctuations and $z$ the extra dimensionality caused by the imaginary time[55]. The continuous reduction of gap anisotropy and its power-law dependence as a function of composition close to $x_c$ together point to a nematic QCP lying beneath the superconducting dome.

Having analyzed the gap anisotropy, we move onto the averaged gap amplitude at $T = 0$ K extracted from the fits. A critical finding is that both $\Delta_0$ and the gap ratio $2\Delta_0/k_B T_c$ are the largest at $x_c$ (the bottom panels of Fig. S7a and Fig. 3c). This fact clearly demonstrates that the electron pairing is strengthened upon approaching the nematic QCP in both the orthogonal and tetragonal phases. The gap ratio increases with Te doping in the nematic phase, but changes only slightly from 4.5 for $x_c < x \leq 0.312$ to 3.85 for $x > 0.312$ in the tetragonal phase (Fig. 3c, blue dashed line). Note that for the anisotropic pairing gap in the nematic phase, although the values of $2\Delta_0/k_B T_c$ are less than the expected value of 3.52 for weak-coupling $s$-wave BCS superconductors, the ratios of gap maximum to $T_c$, i.e., $2\Delta_{max}/k_B T_c$, ranges from 3.2 to 5.4, which are close to or larger than 3.52. For all the compositions in the tetragonal phase, the gap ratio is a bit larger than the BCS value (Fig. 3c, the purple dashed line). This suggest that the superconductivity of FST locates in the strong-coupling regime, in line with a recent upper critical field study[56].

**Doping dependence of the superfluid density at low temperature**

In order to learn about the phase coherence of the superconducting condensate around the nematic QCP, we now turn to the doping dependence of the superfluid density at the base temperature of 5 K [$\Lambda_p^{-1}(5\ K)$]. Zero-temperature superfluid density encodes the phase rigidity of the superconducting order parameter without thermal effects, which is regarded as another key ingredient affecting $T_c$ in unconventional superconductors[57–59]. Nevertheless, replacing the measured $\Lambda_p^{-1}(5\ K)$ with $\Lambda_p^{-1}(0\ K)$, which is obtained by fitting the $\Lambda_p^{-1}(T)$ data to Eq. (1), do not alter its quantitative correlation with $T_c$ (Supplementary Section 6).

In the orthogonal phase, the value of $\Lambda_p^{-1}$ rises with Te doping and becomes relatively saturated close to $x_c$ (Fig. 2d, solid circles). As the doping increased across $x_c$, $\Lambda_p^{-1}$ increases with an even steeper rate. It is noteworthy that such a smooth variation of $\Lambda_p^{-1}$ near $x_c$ is in stark contrast to the sharp dip of low-temperature superfluid density accompanying the AFM QCP in iron pnictide superconductors $BaFe_2(As_{1-x}P_x)_2$ (Fig. 2d, green circles)[15] and $Ba(Fe_{1-x}Co_x)_2As_2$ (Fig. 2d, brown circles)[16].

We find that the doping at which $\Lambda_p^{-1}(T)$ is maximum, denoted as $x_p$, is close to $x_c \sim 0.23$ near $T_c$, but shifts toward $x \sim 0.312$ at low temperatures (Figs. 4a and 4b). It was suggested that the nematic critical fluctuations associated with the QCP may be pinned to nanoscale nematic puddles in the tetragonal phase, inside which the superconducting phase coherence is suppressed while the gap amplitude remains almost unchanged[60]. Within this framework, the low-temperature value of $\Lambda_p^{-1}$ will keep growing beyond $x_c$, until the pinned nematic critical fluctuations completely vanish around $x_p$. At high temperatures close to $T_c$, the influence of nematic critical fluctuations becomes relatively weak owing to the thermal excitations, which can qualitatively explain the temperature evolution of $x_p$. In addition, ARPES measurements on FST/CaF$_2$ films reveal the close proximity of the $d_{xy}$ band to the Fermi energy ($E_F$) for $x > x_c$, leading to an increase of density of state at $E_F$[61]. This may also contribute to the shift of $x_p$ towards large Te doping at low temperatures.

The quantitative relationships between $\Lambda_p^{-1}$ and $T_c$ are distinctly different on the two sides of the nematic QCP. Since $T_c$ reaches maximum around $x_c$ (Fig. 4b, blue squares), the shift of $x_p$ at low temperatures results in a weak correlation between $T_c$ and superfluid density in the tetragonal phase. Particularly, $T_c$ is found to scale linearly with $\Lambda_p^{-1}$ for $x < x_c$ (Fig. 4c, red points and red dashed line). Upon crossing the QCP (Fig. 4c, orange points), the line immediately turns horizontal such that $T_c$ maintains its maximum value for $x_c < x < 0.312$, where $\Lambda_p^{-1}$ increases from 0.076 μm$^{-1}$ to 0.099 μm$^{-1}$ (light blue points in Fig. 4c and Fig. S9). Upon increasing $x$ further, $\Lambda_p^{-1}$ vs. $T_c$ turns backdown in a linear form with a much smaller slope than that of the nematic side. Overall, the SC2 phase exhibits a 'Boomerang' pattern, frequently observed in overdoped cuprate superconductors such as La$_{2-x}$Sr$_x$CuO$_4$[62], Tl$_2$Ba$_2$CuO$_{6+\delta}$[63,64], and Yb$_{0.7}$Ca$_{0.3}$Ba$_{1.6}$Sr$_{0.4}$Cu$_3$O$_{7-\delta}$ [65].

**Discussions**

It is expected that quantum critical fluctuations strengthening the pairing interaction can also lead to strong renormalization to the quasiparticle mass $m^*$ by enhancing electron correlations. For iron pnictide superconductors, several probes indicate a diverging $m^*$ toward the AFM QCP where $T_c$ is promoted[14]. In FST, however, $\Lambda_p^{-1}$ evolves smoothly across the nematic phase boundary, without any notable decrease at $x_c$ (Fig. 2d). The absence of $m^*$ enhancement in the vicinity of a pure nematic QCP suggests the divergence of $m^*$ in the iron pnictides as a consequence of spin fluctuations. While disorder effects may smear out the anticipated sharp dip of $\Lambda_p^{-1}$ at the QCP[15], the absence of $m^*$ enhancement in FST was also reported by a recent upper critical field study[56]. This is also consistent with earlier ARPES studies showing no sign of $m^*$ enhancement near QCP in all orbitals located around the Γ point[66,67]. Such enhancement of pairing strength without affecting $m^*$ suggests that the nematic fluctuations drive Cooper pairing in a different manner compared to the AFM fluctuations.

The linear dependence of $T_c$ on $\Lambda_p^{-1}$ and non-uniform gap ratio (Fig. 3c, red triangles)

in the nematic phase are both beyond the Bardeen-Cooper-Schrieffer paradigm, where $T_c$ is determined by the pairing potential. The former is reminiscent of the Uemura law of underdoped cuprates[58], which was attributed to strong superconducting phase fluctuations owing to small zero-temperature superfluid density[59]. Thus, the observed linear scaling between $T_c$ and superfluid density for $x < x_c$ may also imply strong phase fluctuations in SC1, in line with the giant superconducting fluctuations unveiled in the nematic FeSe[17,68]. Previous DFT calculations and experimental studies of FST have shown that the appearance of nematic order can suppress the normal-state carrier density and thereby reduce the zero-temperature superfluid density[33,69]. In this regard, the potential significant phase fluctuations in SC1 may be at least partially related to the nematic order. In contrast, the $T_c$ of SC2 is governed by the (isotropic) gap amplitude rather than the superfluid density, suggesting that the influence of phase fluctuations is limited without the nematic order. Moreover, both $\Lambda_p^{-1}$ and $\Delta_0$, which are the controlling factors of $T_c$ for SC1 and SC2, respectively, become maximum when approaching the nematic QCP. These results suggest the nematic QCP in FST separates two superconducting phases with distinct pairing symmetries and critical behaviors. Phase fluctuations may play different roles for order parameters with and without anisotropy.

At last, we compare our case with the cuprates. For cuprate superconductors, recent elastoresistivity studies reveal signatures of a nematic QCP near the end point of the pseudogap phase[7–9]. Particularly, the Uemura law applies to the pseudogap regime where a vestigial nematic order has been found[7]. Our finding for the SC1 of FST seems similar to this result. Although a linear scaling between $T_c$ and superfluid density was found in overdoped thin films without nematicity[57], whether it reflects significant phase fluctuations has remained actively debated, mainly because it may be understood under the context of BCS theory for dirty superconductors[70–72]. The 'Boomerang' shaped path of the $T_c$ versus superfluid density for several overdoped cuprates[62–65] is also akin to what we found in SC2. These similarities consistently hint at a critical role played by the nematic QCP on the superconductivity of these two families of high-$T_c$ superconductors. Further efforts are needed to pin down the nematic QCP in cuprate superconductors.

**Conclusion**

In summary, we have employed scanning SQUID susceptometry to map out the superfluid density in unprecedented fine steps across a nematic quantum critical point in the superconducting dome of composition-spread FeSe$_{1-x}$Te$_x$ films. In the nematic phase, the continuous reduction of gap anisotropy and its power-law dependence as a function of composition close to $x_c$ both indicate a pure nematic QCP lying beneath the superconducting dome. The enhanced gap amplitude at $x_c$ further supports the scenario that pairing interaction is strengthened by the quantum fluctuations associated with the nematic QCP. Such a pure nematic QCP divides the two superconducting states, where $T_c$ is dominated by the phase stiffness and the pairing potential for $x < x_c$ and $x > x_c$, respectively. Our results represent the first direct evidence for a pure nematic QCP in the superconducting dome of high $T_c$ superconductors and our methodology paves the way for future investigation of Cooper pairing via nematic quantum fluctuations.

**Acknowledgements**

We would like to acknowledge support by National Key R&D Program of China (Grant No. 2021YFA1400100), Shanghai Municipal Science and Technology Major Project (Grant No. 2019SHZDZX01), National Natural Science Foundation of China (12225412), CAS Project for Young Scientists in Basic Research (2022YSBR-048). The authors are grateful for the stimulating discussions with Steven Kivelson, Yang Qi, Andrey Chubukov and Ziqiang Wang.

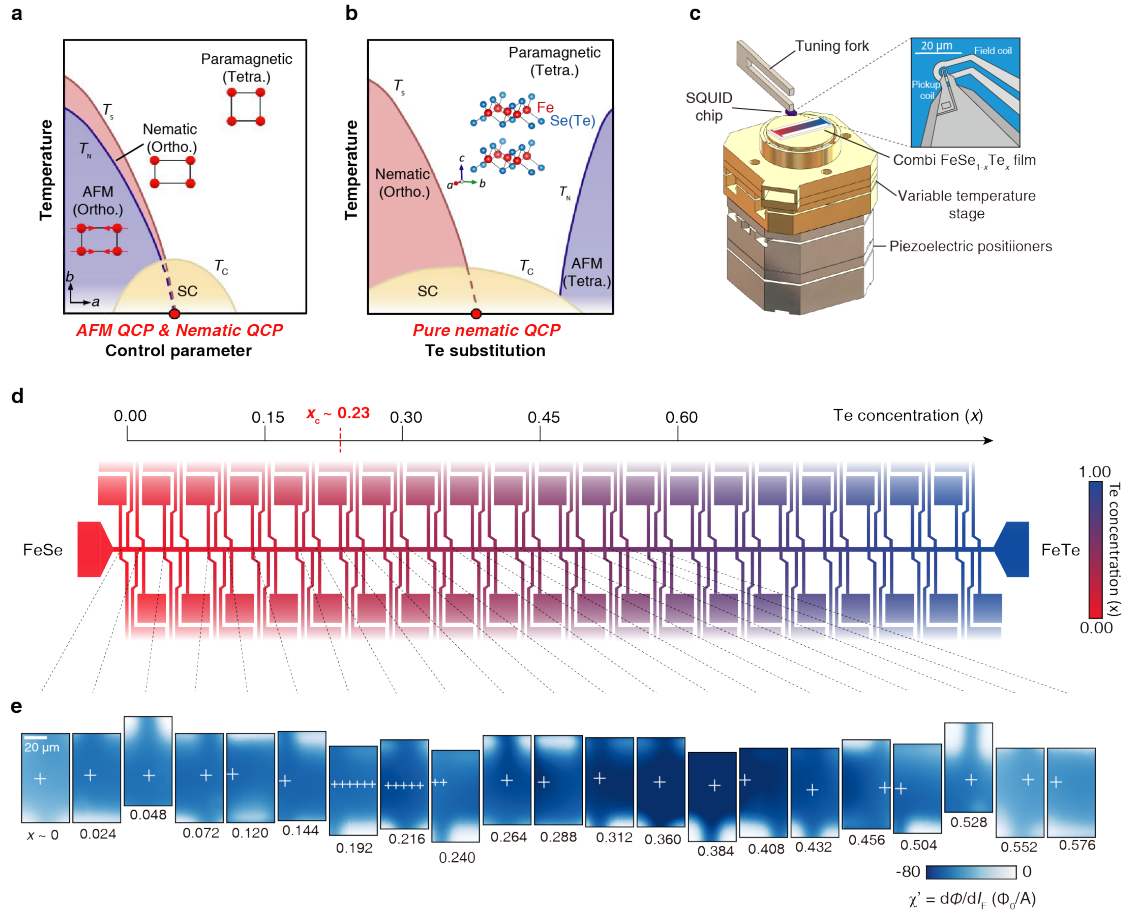

**Fig. 1 | Phase diagram of iron-based superconductors and local susceptometry measurements on the composition-spread FeSe$_{1-x}$Te$_x$ film. a,** General temperature vs nonthermal control parameter phase diagram of iron-based superconductors, demonstrating the close coupling between electronic nematic order and long-range magnetism. $T_s$, $T_N$, and $T_c$ refers to the onset temperature of the nematic phase, the stripe-type antiferromagnetic (AFM) phase, and the superconducting (SC) phase, respectively. **b,** Schematic temperature vs. Te doping phase diagram of FeSe$_{1-x}$Te$_x$. The nematic phase is decoupled with any long-range magnetism at ambient pressure. The inset shows the crystal structure of FeSe$_{1-x}$Te$_x$. 'Ortho.': orthogonal phase; 'Tetra.': tetragonal phase; 'QCP': quantum critical point. **c,** Schematic of the local susceptometry measurements. The inset depicts the nano-SQUID chip with a closeup of the tip, highlighting the field coil and the pickup coil. **d,** Schematic illustration of the composition-spread FeSe$_{1-x}$Te$_x$/CaF$_2$ film, which is patterned into 120 microbridges. The color gradient (red to blue) indicates the continuous variation of Te content ($x$ = 0 to 1), covering the critical Te content at which nematicity vanishes ($x_c \sim 0.23$). **e,** Typical in-phase ac susceptibility images of the film, acquired at 5 K. The distance between the nano-SQUID chip and the film surface is 1 μm. The color bar indicates the strength of diamagnetic susceptibility in units of $\Phi_0/A$, where $\Phi_0 = h/2e$ is the flux quantum. Markers indicate the positions at which the susceptibility approach data are measured.

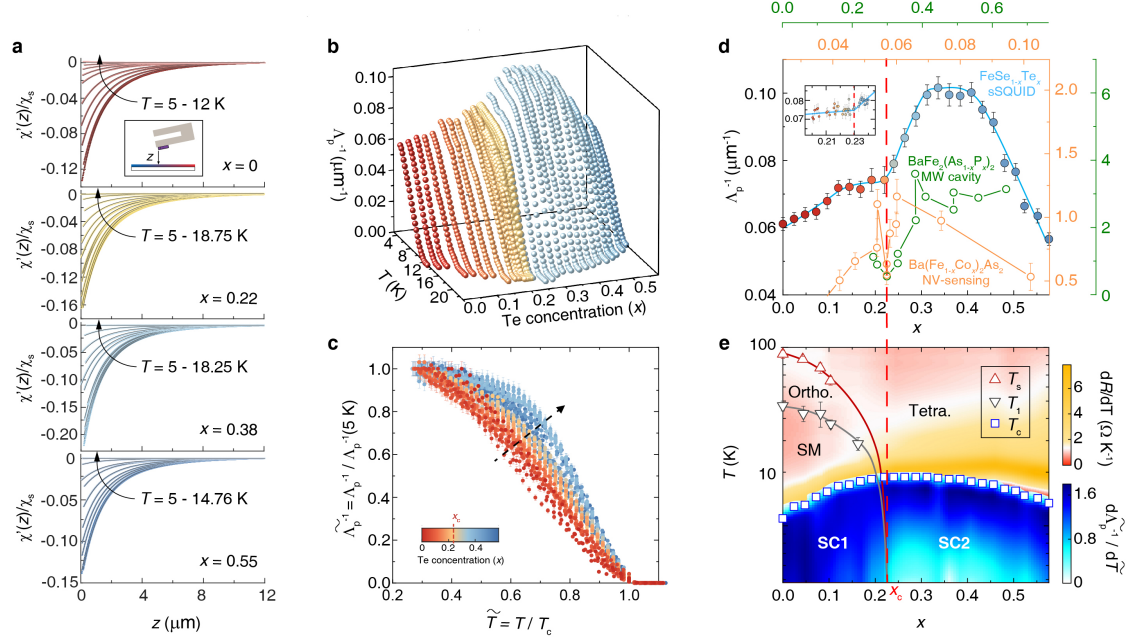

**Fig. 2 | Mapping of the superfluid density as a function of temperature and doping. a,** Normalized in-phase susceptibility $\chi'(z)/\chi_s$ with $\chi_s$ the SQUID self-susceptibility as a function of tip-sample distance for selected compositions ($x$ = 0, 0.23, 0.38, 0.55). The measurement scheme is shown in the inset. The superfluid density ($\propto \Lambda_p^{-1}$) is obtained by fitting the data to the thin diamagnet model (the gray solid lines). **b,** $\Lambda_p^{-1}(T)$ data for 58 compositions, extracted from the temperature-dependent susceptibility approach cures in Fig. S3. **c,** Normalized superfluid density $\widetilde{\Lambda_p^{-1}} = \Lambda_p^{-1}/\Lambda_p^{-1}(5\,K)$ versus reduced temperature $\widetilde{T} = T/T_c$. **d,** Doping dependence of $\Lambda_p^{-1}$ (5 K). The inset shows the enlarged view of the data near $x_c \sim 0.23$. Although a sudden change of $d\Lambda_p^{-1}(5\,K)/dx$ is observed near $x_c$, the value of $\Lambda_p^{-1}$ (5 K) varies smoothly across $x_c$. This is in contrast to the sharp dip of low-temperature $\Lambda_p^{-1}$ at the QCP of the iron pnictide superconductors $BaFe_2(As_{1-x}P_x)_2$ (green circles, Ref.[15]) and $Ba(Fe_{1-x}Co_x)_2As_2$ (brown circles, Ref.[16]). **e,** Phase diagram of the composition-spread $FeSe_{1-x}Te_x$ film. The temperature derivative of resistance, $dR/dT$, and $d\widetilde{\Lambda_p^{-1}}/d\widetilde{T}$ are shown in two different colormaps above and below $T_c$, respectively. The nematic transition temperature $T_s$ (red triangles) is determined by the peak of $dR/dT$ (Supplementary Section 4). $T_1$ (black triangles) is the upper bound temperature of the strange-metal state (Supplementary Section 5). The superconducting transition temperature $T_c$ (blue squares) is determined by the onset of diamagnetism. 'SM': strange metal. An abrupt change of $d\widetilde{\Lambda_p^{-1}}/d\widetilde{T}$ across $x_c$ suggests two distinct superconducting state: SC1 for $x < x_c$ and SC2 for $x > x_c$.

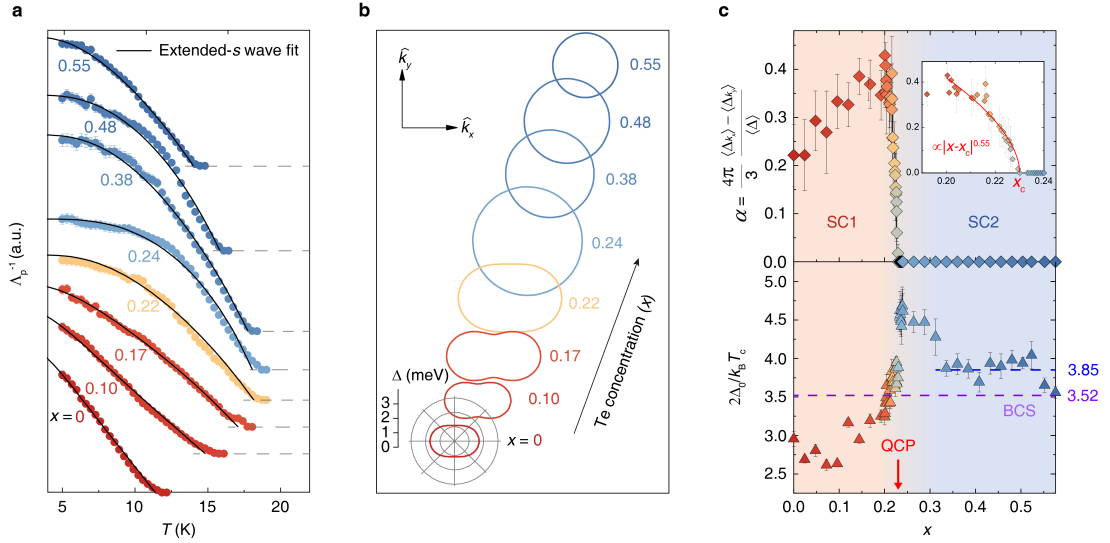

**Fig. 3 | Evolution of the pairing gap symmetry across the nematic QCP. a,** Temperature-dependent superfluid density $[\Lambda_p^{-1}(T)]$ for selected doping levels. The black lines are the best fits of Eq. (1) with the extended $s$-wave gap. The corresponding momentum dependences of the gaps at $T = 0$ K are shown in **b**. The gap structure exhibits a transition from the $C_2$ symmetry to the isotropic symmetry near $x_c \sim 0.23$. **c,** Summary of the fitting parameters. Top panel: $x$ dependence of the anisotropy parameter $\alpha$, which reflects the difference between the gap values along $k_x$ and $k_y$ directions $[\alpha = \frac{4\pi}{3}\frac{\langle\Delta_{k_x}\rangle - \langle\Delta_{k_y}\rangle}{\langle\Delta\rangle}]$. The inset depicts the power-law dependence of $\alpha$ as a function of $x$ close to $x_c$, i.e., $\alpha \propto |x - x_c|^\beta$ with $\beta = 0.55 \pm 0.11$ (the red line). Bottom panel: $x$ dependence of the averaged gap value at zero-temperature $\Delta_0 = \langle\Delta(T = 0\text{ K})\rangle$ divided by its critical temperature $T_c$. The weak-coupling BCS theory predicts $2\Delta_0/k_B T_c = 3.52$ (The purple dashed line).

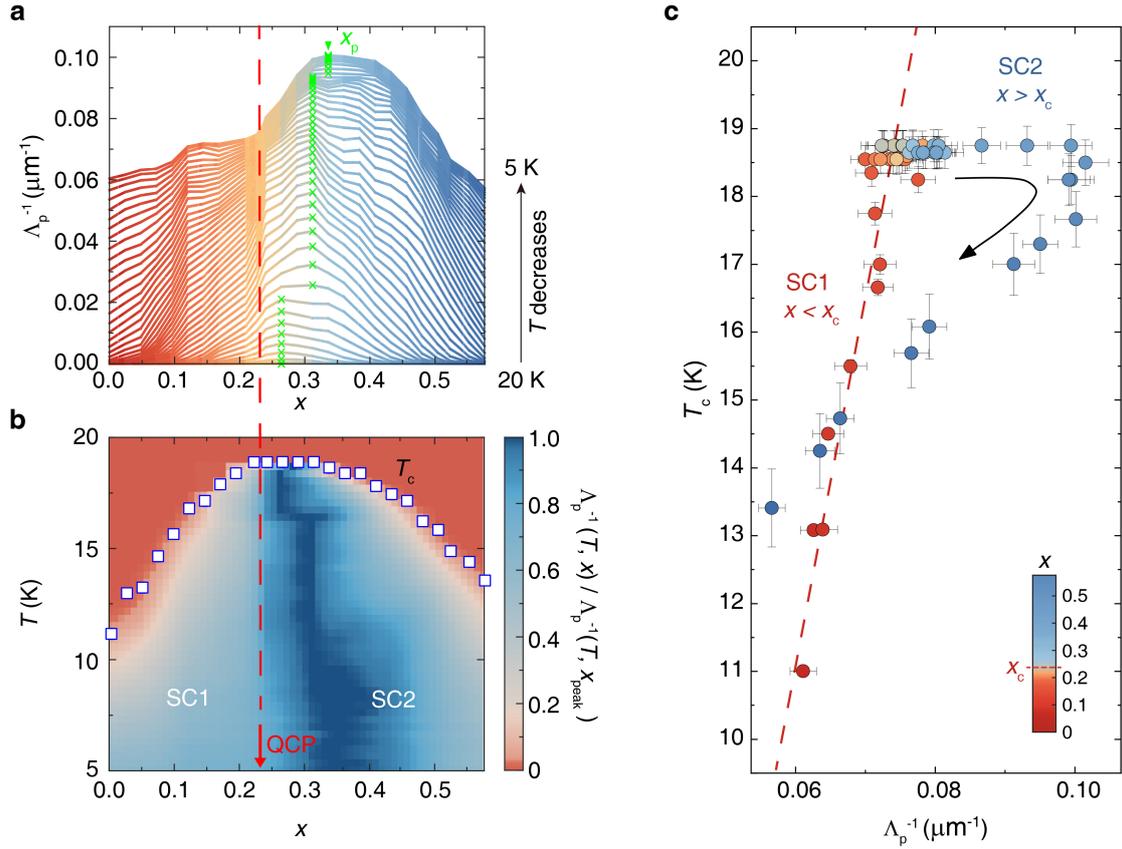

**Fig. 4 | Relationship between $T_c$ and low-temperature superfluid density. a,** Superfluid density ($\propto \Lambda_p^{-1}$) as a function of Te content ($x$) at different temperatures from 20 K to 5 K (the arrow indicates decreasing temperature). The doping at which $\Lambda_p^{-1}$ peaks is denoted by $x_p$ (the green crosses), which deviates from $x_c$ and moves toward higher Te content as temperature decreases. **b,** Color plot of normalized superfluid density $\Lambda_p^{-1}(T,x)/\Lambda_p^{-1}(T,x_p)$. The blue squares represent the superconducting transition temperature $T_c$. **c,** $T_c$ versus $\Lambda_p^{-1}$ (5 K) for different compositions. The Te content is marked by its color (see the color bar). The arrow points toward increasing Te doping. Their relation exhibits a 'Boomerang'-like structure with the QCP at the first turning point.